\documentclass[%
 reprint,
 amsmath,amssymb,
 aps,
]{revtex4-2}
\usepackage[most]{tcolorbox}
\usepackage{graphicx}
\usepackage{dcolumn}
\usepackage{bm}
\usepackage{hyperref}
\usepackage{slashed}
\usepackage{subfloat}
\usepackage{float}
\usepackage{pgfplots}
\usepackage{subfigure}
\usepackage{etoolbox}   
\usepackage{orcidlink}
\usepackage{geometry}
\usepackage{mathrsfs}
\definecolor{revUnified}{RGB}{190,30,45} 

\begin{document}

\fontsize{7.6}{8.6}\selectfont

\title{Charged Klein--Gordon modes on an Ellis wormhole: magnetic confinement and conditional Heun-polynomial solvability}

\author{Abdullah Guvendi
}
\email{abdullah.guvendi@erzurum.edu.tr}
\affiliation{Department of Basic Sciences, Erzurum Technical University, 25050, Erzurum, Türkiye}

\author{Omar Mustafa
}
\email{omar.mustafa@emu.edu.tr (Corresponding Author)}
\affiliation{Department of Physics, Eastern Mediterranean University, 99628, G. Magusa, North Cyprus, Mersin 10 - Türkiye}

\date{\today}

\begin{abstract}
{\fontsize{7.6}{8.6}\selectfont \setlength{\parindent}{0pt}

We study a charged Klein--Gordon field on the complete $(2+1)$-dimensional Ellis wormhole, coupled to the pullback of a uniform ambient magnetic field. The induced gauge potential is regular, whereas the intrinsic magnetic pseudoscalar reverses sign across the throat. For $\alpha_B>0$, the radial operator is unitarily equivalent to a self-adjoint one-dimensional Schr\"odinger operator with a smooth confining potential and discrete spectrum. The radial equation is reduced to confluent-Heun form, and polynomial solvability is shown to require two independent termination conditions. These conditions yield degree-zero Gaussian modes, degree-one states of both global parities, and representative degree-two even states. Exact Ellis-measure normalization is obtained for the degree-zero and degree-one modes. An independent fourth-order full-line finite-difference calculation validates the analytic benchmarks and determines the first eight eigenvalues and their parities. In the double-well regime, the forbidden-region action determines the leading semiclassical exponential of the even--odd level splitting.
}
\end{abstract}


\maketitle

\section{Introduction}
\label{sec:introduction}

\setlength{\parindent}{0pt}

Quantum fields on curved backgrounds provide a framework for determining how geometry modifies relativistic wave dynamics. In curved spacetime, the spectral and localization properties of a field are determined not only by its mass and quantum numbers but also by the geometry of the underlying manifold and, for charged fields, by its coupling to an electromagnetic gauge connection \cite{parker1,a,b,c}. The simultaneous presence of curvature and gauge coupling can therefore generate physical structures. Wormhole geometries provide a particularly useful setting for such investigations because they combine a nontrivial global topology with a regular throat connecting distinct asymptotic regions.

The geometric concept of a bridge between distinct spacetime regions was introduced by Einstein and Rosen \cite{1}, while the Morris-Thorne framework established a general description of traversable wormholes characterized by a regular throat and finite redshift function \cite{2}. Since then, wormhole geometries have been investigated in a wide range of gravitational theories and spacetime dimensions, including three-dimensional gravity, modified theories of gravity, and quantum-field systems propagating on wormhole backgrounds \cite{3,4,6,7,8,9,10,11,12,13,b,13b,13c,13d,14a,14b,14c,17,17a}. These studies have examined how topology, curvature, and global geometry affect classical and quantum fields.

For a static and circularly symmetric $(2+1)$-dimensional traversable wormhole, the geometry can be written in Morris--Thorne-type coordinates as \cite{a,8}
\begin{equation}
ds^{2}
=
-e^{2\Phi(r)}dt^{2}
+
\frac{dr^{2}}{1-b(r)/r}
+
r^{2}d\phi^{2},
\qquad
\phi\sim\phi+2\pi,
\label{eq:intro_wormhole_metric}
\end{equation}
where $\Phi(r)$ is the redshift function and $b(r)$ is the shape function. The throat is located at the minimum circumferential radius, while the two asymptotic regions are represented by the two branches of the radial geometry. This form provides the general static, circularly symmetric setting for the Ellis geometry considered below. {In the zero-redshift Ellis case, $\Phi(r)=0$ and $b(r)=b_0^2/r$; the proper radial coordinate satisfies $dx=dr/\sqrt{1-b_0^2/r^2}$ on each branch.}

{The Ellis wormhole provides a regular realization} \cite{Ellis}. Its spatial section possesses a finite minimal circle and two asymptotically distinct ends, while its intrinsic two-dimensional geometry is isometric to a catenoid \cite{Ellis,25,25a,25b,25c}. This correspondence provides an explicit geometrical representation of the wormhole surface in an auxiliary Euclidean space without introducing the embedding dimension as an additional physical coordinate. The catenoidal realization is especially useful because catenoid geometries have independently been studied as curved backgrounds for geometry-induced quantum effects, relativistic wave propagation, and electronic systems \cite{a,13d,25,25a,25b,25c}.

{A complementary recent analysis of the same Ellis background derived exact scalarization thresholds and onset laws for a scalar field nonminimally coupled to curvature \cite{GuvendiScalarization2026}; the present work instead concerns magnetic confinement of a charged Klein--Gordon field.}

{The embedding representation provides a construction of an external electromagnetic field. We define a uniform magnetic field in the auxiliary Euclidean space and pull back its gauge connection to the physical surface} \cite{a1,a2}. The auxiliary space then serves only as a geometrical device for specifying the external field, whereas the field dynamics remains confined to the physical wormhole manifold. The resulting intrinsic gauge potential is determined by the geometry of the embedded surface and therefore need not correspond to a uniform magnetic scalar with respect to the intrinsic metric (see also \cite{a1,a2}).

{The present analysis concerns a charged Klein--Gordon field on the complete Ellis surface, with the external connection fixed by the pullback of a uniform ambient magnetic field. It establishes the self-adjoint radial operator on the full line, the two-condition classification of Heun-polynomial modes in both global parity sectors, exact normalization of the degree-zero and degree-one modes with the Ellis measure, and the placement of the analytic states within the numerically determined confined spectrum.}

In this work, we investigate charged Klein-Gordon modes on the $(2+1)$-dimensional Ellis wormhole in the presence of a magnetic field defined uniformly in the auxiliary Euclidean space associated with its catenoidal realization. The paper is organized as follows. Section~\ref{sec:ellis_geometry} develops the Ellis wormhole geometry, its catenoidal embedding, the global proper radial coordinate, and the intrinsic curvature. Section~\ref{sec:ellis_gauge} constructs the uniform magnetic field in the auxiliary Euclidean space and derives its gauge pullback and intrinsic electromagnetic field. Section~\ref{sec:charged_KG} formulates the charged Klein-Gordon dynamics and obtains the radial and one-dimensional representations. Section~\ref{sec:radial_reduction} analyzes the radial equation within the confluent-Heun framework and determines the polynomial truncation conditions and associated spectrum. {The analysis also solves both global parity sectors, treats the degree-zero and degree-one polynomial families, and tests analytic states against a full-line finite-difference spectrum; a separate subsection gives exact radial normalization and localization densities.} Section~\ref{sec:conclusion} summarizes the principal results and their physical implications.

\section{Wormhole Geometry}
\label{sec:ellis_geometry}

\setlength{\parindent}{0pt}

\begin{figure}[!t]
\centering
\includegraphics[width=0.8\linewidth]{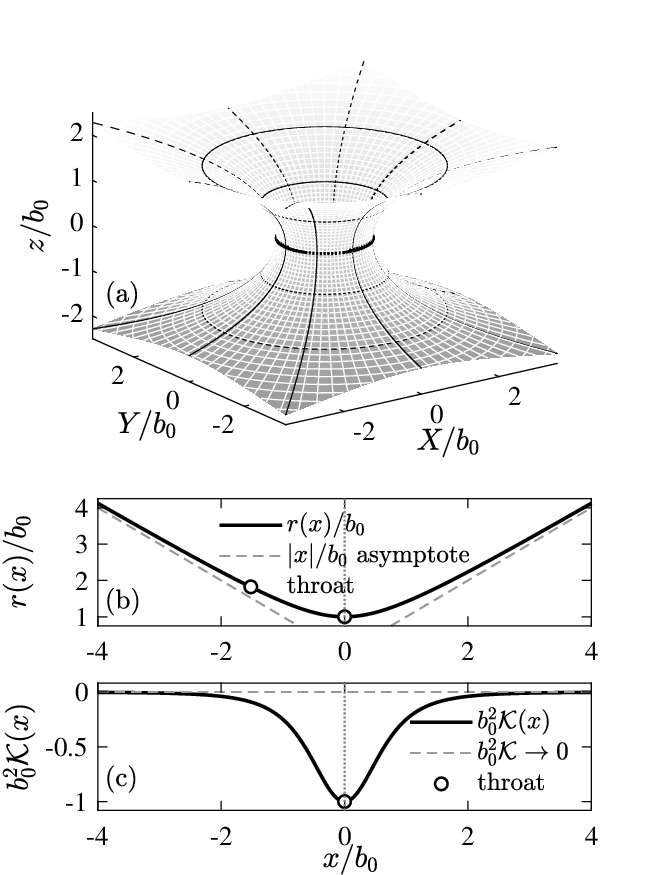}
\caption{\fontsize{7.6}{8.6}\selectfont
Wormhole geometry. (a) Isometric embedding of the wormhole surface, Eq.~\eqref{eq:2.1}, which corresponds to a catenoid in an auxiliary Euclidean space; the throat at \(z=0\) has radius \(\rho_{\rm th}=b_0\), with \(b_0>0\).
(b) Intrinsic radius \(r(x)/b_0=\sqrt{1+(x/b_0)^2}\), where \(x=b_0\sinh(z/b_0)\); the dashed curve denotes \(r/b_0\sim|x|/b_0\) and the marker indicates \(x=0\).
(c) Dimensionless Gaussian curvature \(b_0^2\mathcal{K}(x)=-[1+(x/b_0)^2]^{-2}\), with \(b_0^2\mathcal{K}(0)=-1\) and \(\mathcal{K}\to0^{-}\) as \(|x|\to\infty\). Here \(\phi\in[0,2\pi)\) and \(x\in(-\infty,\infty)\); all lengths are normalized by \(b_0\).}
\label{fig:1}
\end{figure}

We consider the equatorial spatial geometry of the Ellis wormhole, which provides a regular two-dimensional wormhole surface with a finite throat connecting two asymptotic ends. For the massless, zero-redshift configuration, this geometry admits an exact isometric embedding into an auxiliary Euclidean space. Introducing cylindrical coordinates in the embedding space, the surface is parametrized by
\begin{equation}
\mathbf{R}(z,\phi)=
\left(
b_0\cosh\frac{z}{b_0}\cos\phi,\,
b_0\cosh\frac{z}{b_0}\sin\phi,\,
z
\right),
\qquad
\phi\sim\phi+2\pi,
\label{eq:2.1}
\end{equation}
where \(z\) is the auxiliary axial coordinate and \(b_0>0\) is the throat radius. The corresponding cylindrical radius is
\begin{equation}
\rho(z)
=
b_0\cosh\frac{z}{b_0},
\label{eq:2.2}
\end{equation}
whose minimum value is \(b_0\) at \(z=0\). Thus, the embedded surface is a catenoid, with the circle \(z=0\) representing its minimal throat. The embedding space is used solely as an auxiliary Euclidean realization of the intrinsic geometry and does not introduce an additional physical dimension into the spacetime. The catenoidal embedding and the throat are shown in Fig.~\ref{fig:1}(a). The induced metric is obtained by pulling back the Euclidean line element,
\begin{equation}
d\ell^2
=
dX^2+dY^2+dZ^2
=
\cosh^2\frac{z}{b_0}
\left(
dz^2+b_0^2d\phi^2
\right).
\label{eq:2.3}
\end{equation}

It is convenient to introduce the proper radial coordinate
\begin{equation}
x=b_0\sinh\frac{z}{b_0},
\qquad
z=b_0\operatorname{arsinh}\frac{x}{b_0},
\label{eq:2.4}
\end{equation}
for which
\begin{equation}
dx=
\cosh\frac{z}{b_0}\,dz,
\qquad
x^2+b_0^2=b_0^2\cosh^2\frac{z}{b_0}.
\label{eq:2.5}
\end{equation}
The intrinsic metric therefore becomes
\begin{equation}
d\ell^2=dx^2+r^2(x)d\phi^2,\qquad r(x)=\sqrt{x^2+b_0^2}.
\label{eq:2.6}
\end{equation}
Here \(x\in(-\infty,\infty)\) is a global proper radial coordinate, while \(r(x)\) is the circumferential radius at fixed \(x\). At the throat,
\begin{equation}
r(0)=b_0,
\qquad
r'(0)=0,
\label{eq:2.7}
\end{equation}
so that \(x=0\) is a smooth minimal circle of the intrinsic geometry. The corresponding circumferential-radius profile is shown in Fig.~\ref{fig:1}(b).

For the metric in Eq.~\eqref{eq:2.6}, the Gaussian curvature is \cite{27,17,17a}
\begin{equation}
\mathcal{K}(x)
=
-\frac{r''(x)}{r(x)}
=
-\frac{b_0^2}
{\left(x^2+b_0^2\right)^2}.
\label{eq:2.8}
\end{equation}
It is finite throughout the surface, with
\begin{equation}
\mathcal{K}(0)
=
-\frac{1}{b_0^2}
\label{eq:2.9}
\end{equation}
at the throat. The resulting curvature profile and its asymptotic decay are shown in Fig.~\ref{fig:1}(c). The throat is regular, and the Gaussian curvature remains finite there. The negative Gaussian curvature is consistent with the catenoidal embedding, which is a minimal surface with vanishing mean curvature and principal curvatures of opposite signs \cite{25b,25c}.

\section{Uniform Magnetic Field and Its Intrinsic Gauge Pullback}
\label{sec:ellis_gauge}

\setlength{\parindent}{0pt}

We next introduce a uniform magnetic field in the auxiliary Euclidean embedding space and determine the electromagnetic potential induced on the Ellis surface by pullback. Let the ambient magnetic field be oriented along the embedding-space $Z$-axis,
\begin{equation}
\boldsymbol{\mathcal B}_0=\mathcal B_0\,\hat{\mathbf Z}.
\label{eq:2.13}
\end{equation}
A rotationally symmetric gauge potential representing this uniform field is conveniently written as the one-form
\begin{equation}
A=\frac{\mathcal B_0}{2}\left(X\,dY-Y\,dX\right),
\label{eq:2.14}
\end{equation}
whose exterior derivative gives the corresponding field-strength two-form,
\begin{equation}
dA=\mathcal B_0\,dX\wedge dY.
\label{eq:2.15}
\end{equation}
This gauge is adapted to the axial symmetry of the embedded surface and therefore provides a natural starting point for constructing the intrinsic electromagnetic coupling. Using the cylindrical representation
\begin{equation}
X=\rho(z)\cos\phi,\qquad Y=\rho(z)\sin\phi,
\label{eq:2.16}
\end{equation}
one finds
\begin{equation}
X\,dY-Y\,dX=\rho^2(z)\,d\phi.
\label{eq:2.17}
\end{equation}
The pullback of the ambient gauge potential onto the Ellis surface is consequently
\begin{equation}
\mathcal A=\iota^*A=\frac{\mathcal B_0}{2}\rho^2(z)\,d\phi,\qquad
\rho^2(z)=x^2+b_0^2=r^2(x).
\label{eq:2.18}
\end{equation}
Accordingly, the intrinsic potential takes the form

\begin{figure}[H]
\centering
\includegraphics[width=0.9\linewidth]{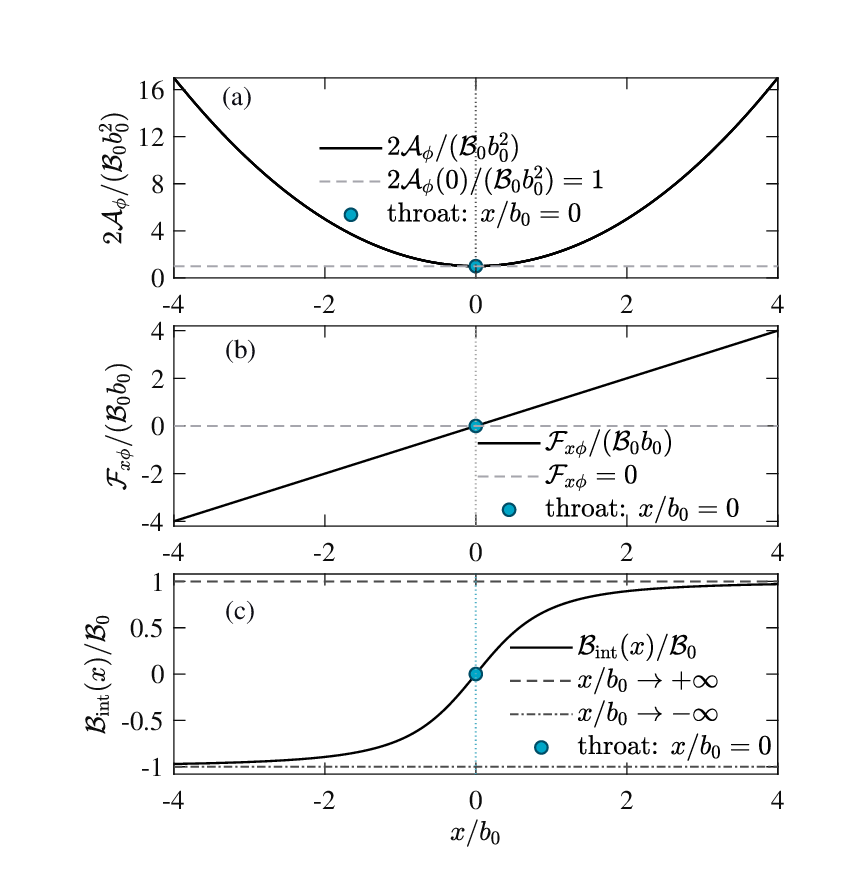}
\caption{\fontsize{7.6}{8.6}\selectfont
Intrinsic magnetic pullback on the Ellis wormhole. (a) Dimensionless azimuthal gauge potential
$2\mathcal A_\phi/(\mathcal B_0b_0^2)=1+(x/b_0)^2$. (b) Dimensionless field-strength component
$\mathcal F_{x\phi}/(\mathcal B_0b_0)=x/b_0$. (c) Intrinsic magnetic scalar
$\mathcal B_{\rm int}/\mathcal B_0=(x/b_0)/\sqrt{1+(x/b_0)^2}$, which vanishes at the throat and approaches $-\!1$ and $+\!1$ at the two asymptotic ends.}
\label{fig:2}
\end{figure}

\begin{equation}
\mathcal A
=
\mathcal A_\phi(x)\,d\phi,
\qquad
\mathcal A_\phi(x)
=
\frac{\mathcal B_0}{2}
\left(
x^2+b_0^2
\right).
\label{eq:2.19}
\end{equation}
The intrinsic azimuthal gauge potential obtained from the pullback is shown in Fig.~\ref{fig:2}(a).

The pulled-back connection has only an azimuthal component, $\mathcal A_\phi=\mathcal B_0r^2/2$. Its radial dependence is fixed by the circumferential radius of the embedded surface; the ambient field remains uniform in the auxiliary Euclidean space. The intrinsic electromagnetic field strength follows from the exterior derivative of the pulled-back potential,
\begin{equation}
\mathcal F
=
d\mathcal A
=
\mathcal B_0 x\,dx\wedge d\phi.
\label{eq:2.20}
\end{equation}
The corresponding field-strength component is displayed in Fig.~\ref{fig:2}(b), showing its linear dependence on the proper radial coordinate and its sign reversal across the throat.

The corresponding intrinsic area two-form is
\begin{equation}
d\Sigma
=
r(x)\,dx\wedge d\phi.
\label{eq:2.21}
\end{equation}
Hence, the oriented intrinsic magnetic pseudoscalar obtained by the Hodge dual of \(\mathcal F\)
\begin{equation}
\mathcal B_{\mathrm{int}}(x)
=
\frac{\mathcal F_{x\phi}}{r(x)}
=
\mathcal B_0
\frac{x}{\sqrt{x^2+b_0^2}}.
\label{eq:2.22}
\end{equation}
The resulting intrinsic magnetic scalar is shown in Fig.~\ref{fig:2}(c); it vanishes at the throat and approaches opposite asymptotic values on the two ends of the wormhole.

The radial dependence of $\mathcal B_{\mathrm{int}}$ in Eq.~\eqref{eq:2.22} follows from the pullback of the uniform ambient magnetic field onto the curved wormhole surface. More precisely, $\mathcal B_{\mathrm{int}}$ is obtained by contracting the intrinsic field-strength two-form with the oriented area form of the surface, so its local value is fixed by the surface orientation and intrinsic area element. With the continuous orientation determined by the chosen parametrization, the unit normal varies smoothly across the throat, and its \(Z\)-component changes sign between the two asymptotic ends, giving the sign change of $\mathcal B_{\mathrm{int}}$ across the throat. This reversal is a geometric projection effect and does not indicate any discontinuity of the electromagnetic two-form $\mathcal F$ or the gauge potential $\mathcal A$, both of which remain regular at $x=0$.

{The pullback connection also has a global gauge-invariant characterization through its $U(1)$ Wilson loop (holonomy) around a circle of fixed $x$. For a charged field, the gauge-invariant phase is the exponential of the gauge-covariant circulation of the connection \cite{AharonovBohm1959}. At the throat, where the circular orbit has radius $b_0$, it is
\begin{equation}
\mathcal W_{\rm th}
=\exp\!\left(\frac{iq}{\hbar}\oint_{x=0}\mathcal A\right)
=\exp\!\left(\frac{iq}{\hbar}\,2\pi\mathcal A_\phi(0)\right)
=\exp\!\left(\frac{iq\pi\mathcal B_0b_0^2}{\hbar}\right),
\label{eq:throat_wilson_holonomy}
\end{equation}
where the last equality follows directly from the pulled-back potential in Eq.~\eqref{eq:2.19}. The phase is invariant under single-valued $U(1)$ gauge transformations and is defined modulo $2\pi$. It is generally nontrivial, although it becomes unity when $q\mathcal B_0b_0^2/(2\hbar)$ is an integer. This global holonomy is distinct from the local field scalar: $\mathcal B_{\rm int}(0)=0$ states that the oriented magnetic density vanishes at the throat, not that the connection has zero circulation around the throat circle. In the remainder of the paper we use $\hbar=1$.}
{The ambient field is an externally prescribed background. The electromagnetic field is not determined self-consistently by Maxwell equations on the wormhole, and the auxiliary embedding coordinates do not enter the field dynamics. The model describes charged scalar excitations coupled to a specified external connection on the curved surface.}

\section{Charged Klein-Gordon Dynamics}
\label{sec:charged_KG}

\setlength{\parindent}{0pt}

\begin{figure}[t]
\centering
\includegraphics[width=0.8\linewidth]{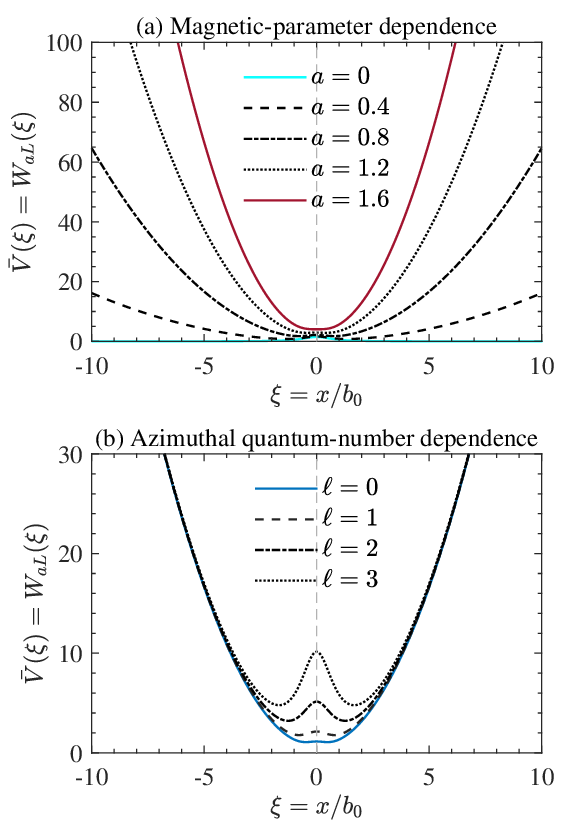}
\caption{\fontsize{7.6}{8.6}\selectfont
Dimensionless effective potential $\bar V(\xi)=b_0^2V_{\rm eff}(b_0\xi)$ for charged Klein-Gordon modes on the Ellis wormhole, with the dimensionless variables defined in Eq.~\eqref{eq:dimensionless_parameters}. (a) Dependence on the dimensionless magnetic-field strength $a$ for fixed azimuthal quantum number $\ell=1$, with $a=0,\,0.4,\,0.8,\,1.2,$ and $1.6$. Increasing $a$ strengthens the confining quadratic contribution $a^2(1+\xi^2)$. (b) Dependence on the azimuthal quantum number for fixed $a=0.8$, with $\ell=0,\,1,\,2,$ and $3$. In both panels, the plotted potential is the exact dimensionless form of Eq.~\eqref{eq:2.40},
$\bar V(\xi)=a^2(1+\xi^2)+\ell^2/(1+\xi^2)+(2-\xi^2)/[4(1+\xi^2)^2]$.
The potential is regular at the throat $\xi=0$ and is even under $\xi\rightarrow-\xi$. For $a>0$, $\bar V(\xi)\sim a^2\xi^2$ as $|\xi|\rightarrow\infty$, giving magnetic confinement, whereas the zero-field case $a=0$ approaches zero at both asymptotic ends.}
\label{fig:3}
\end{figure}

We consider a complex scalar field of mass \(m\) and electric charge \(q\)
propagating on the static \((2+1)\)-dimensional Ellis geometry with \(c=1\) and
signature \((- ,+,+)\),
\begin{equation}
ds^{2}=-dt^{2}+dx^{2}+r^{2}(x)d\phi^{2}.
\label{eq:2.25}
\end{equation}
The metric determinant and inverse metric are
\begin{equation}
g_{\mu\nu}
=
\operatorname{diag}
\left(
-1,\,1,\,r^{2}(x)
\right),
\quad
g^{\mu\nu}
=
\operatorname{diag}
\left(
-1,\,1,\,\frac{1}{r^{2}(x)}
\right),
\end{equation}
with \(\sqrt{-g}=\,r(x)\). The interaction with the electromagnetic field is introduced through
minimal coupling, \(D_{\mu}=\partial_{\mu}-\frac{iq}{\hbar}\mathcal{A}_{\mu}\), where \(\mathcal{A}_{\mu}\) is the electromagnetic gauge potential. In units where $c=1=\hbar$, the gauge-covariant Klein-Gordon equation consequently takes the form \cite{omar2026}
\begin{equation}
\left[
\frac{1}{\sqrt{-g}}
D_{\mu}
\left(
\sqrt{-g}\,
g^{\mu\nu}D_{\nu}
\right)
-m^{2}
\right]\Psi
=
0 .
\label{eq:2.28}
\end{equation}
For the intrinsic gauge potential specified in Eq.~\eqref{eq:2.19}, the azimuthal derivative contains the magnetic coupling. With $c=1=\hbar$, the covariant Klein-Gordon equation reduces to
\begin{equation}
\left[
-\partial_{t}^{2}
+
\frac{1}{r}
\frac{d}{dx}
\left(
r\frac{d}{dx}
\right)
+
\frac{1}{r^{2}}
\left(
\partial_{\phi}
-
\frac{iq\mathcal{B}_{0}}{2}r^{2}
\right)^{2}
-
m^{2}
\right]\Psi
\textcolor{revUnified}{= 0.}
\label{eq:2.29}
\end{equation}
The static and axially symmetric background permits simultaneous separation of the temporal and azimuthal variables. We therefore use the mode decomposition
\begin{equation}
\Psi(t,x,\phi)
=
e^{-iEt}
e^{i\ell \phi } 
R_{\ell}(x),
\label{eq:2.30}
\end{equation}
where \(E\) denotes the relativistic energy, \(R_{\ell}(x)\) is the radial mode function, and $\ell$ is the azimuthal quantum number. {Single-valuedness under $\phi\mapsto\phi+2\pi$ requires $\ell\in\mathbb Z$; noninteger values would require a twisted angular boundary condition and are not considered here.} For later convenience, we introduce the non-negative magnetic scale
\begin{equation}
\alpha_{B}
=
\frac{|q\mathcal{B}_{0}|}{2},
\qquad
\alpha_{B}\geq0,
\end{equation}
together with the sign factor
\begin{equation}
\sigma_{B}
=
\operatorname{sgn}(q\mathcal{B}_{0}),
\qquad
q\mathcal{B}_{0}
=
2\sigma_{B}\alpha_{B}.
\label{eq:2.31}
\end{equation}
For \(q\mathcal{B}_{0}\neq0\), one has \(\sigma_{B}=\pm1\). The zero-field case is obtained continuously by setting \(\alpha_{B}=0\). Upon inserting the separated ansatz
\eqref{eq:2.30} into Eq.~\eqref{eq:2.29}, the radial equation therefore takes the form
\begin{equation}
R_{\ell}''
+
\frac{x}{x^{2}+b_{0}^{2}}R_{\ell}'
+
\left[
\Lambda_{\ell}(E)
-
\alpha_{B}^{2}(x^{2}+b_{0}^{2})
-
\frac{\ell^{2}}{x^{2}+b_{0}^{2}}
\right]
R_{\ell}
=
0 ,
\label{eq:2.33}
\end{equation}
where all constants have been collected into the spectral quantity
\begin{equation}
\Lambda_{\ell}(E)=E^{2}-m^{2}+2\sigma_{B}\alpha_{B}\ell .
\label{eq:2.33a}
\end{equation}
Thus, for fixed \(\alpha_{B}\), \(\sigma_{B}\), and \(\ell\), the differential operator governing the radial profile is independent of \(E\), while the relativistic energy enters through the spectral relation \(\Lambda_{\ell}(E)\). The structure of Eq.~\eqref{eq:2.33} is particularly transparent when
the radial derivative is expressed in divergence form. The radial equation is equivalently written as
\begin{equation}
-\frac{1}{r(x)}
\frac{d}{dx}
\left[
r(x)\frac{dR_{\ell}}{dx}
\right]
+
\left[
\alpha_{B}^{2}r^{2}(x)
+
\frac{\ell^{2}}{r^{2}(x)}
\right]
R_{\ell}(x)
=
\Lambda_{\ell}(E)R_{\ell}(x).
\label{eq:2.35}
\end{equation}
This identifies the energy-independent radial differential operator
\begin{equation}
\mathcal{H}_{\ell}
=
-\frac{1}{r(x)}
\frac{d}{dx}
\left[
r(x)\frac{d}{dx}
\right]
+
\alpha_{B}^{2}r^{2}(x)
+
\frac{\ell^{2}}{r^{2}(x)},
\end{equation}
so that the radial spectral problem can be expressed compactly as
\begin{equation}
\mathcal{H}_{\ell}R_{\ell}
=
\Lambda_{\ell}(E)R_{\ell}.
\end{equation}
The divergence form also makes explicit the natural radial measure with
respect to which the differential expression is symmetric. Namely,
the radial Hilbert space is
\begin{equation}
\mathscr{H}_{\rm rad}
=
L^{2}\!\left(
\mathbb{R},r(x)\,dx
\right),
\end{equation}
with inner product
\begin{equation}
\langle R_{1},R_{2}\rangle_{\rm rad}
=
\int_{-\infty}^{\infty}
r(x)\,
R_{1}^{*}(x)R_{2}(x)\,dx .
\label{eq:2.36}
\end{equation}
For \(b_{0}>0\), the weight \(r(x)=\sqrt{x^{2}+b_{0}^{2}}\) is strictly
positive and smooth on the entire real line. Hence no singular
endpoint is introduced by the throat, and the radial problem is
naturally formulated on the complete coordinate domain
\(x\in\mathbb{R}\). It is useful to remove the nontrivial radial weight and the associated
first-derivative term simultaneously. This is accomplished by introducing the rescaled one-dimensional wavefunction
\(u_{\ell}\) through the unitary map
\begin{equation}
u_{\ell}(x)=\sqrt{r(x)}\,R_{\ell}(x),
\qquad
R_{\ell}(x)=\frac{u_{\ell}(x)}{\sqrt{r(x)}}
=\frac{u_{\ell}(x)}{(x^{2}+b_{0}^{2})^{1/4}} .
\label{eq:2.37}
\end{equation}
{Here $R_{\ell}$ is the radial amplitude in the original Klein--Gordon
field, whereas $u_{\ell}$ is the wavefunction in the unweighted
one-dimensional Schr\"odinger representation; they are not independent
radial fields. The map $\mathscr U:R_{\ell}\mapsto u_{\ell}=\sqrt r R_{\ell}$
has inverse $\mathscr U^{-1}:u_{\ell}\mapsto R_{\ell}=u_{\ell}/\sqrt r$.
Indeed, under this transformation, $r(x)|R_{\ell}(x)|^{2}=|u_{\ell}(x)|^{2}$,
and therefore
\begin{equation}
\int_{-\infty}^{\infty}
r(x)\,
\left|R_{\ell}(x)\right|^{2}\,dx
=
\int_{-\infty}^{\infty}
\left|u_{\ell}(x)\right|^{2}\,dx .
\end{equation}
More generally,
\begin{equation}
\langle R_{1},R_{2}\rangle_{\rm rad}
=
\int_{-\infty}^{\infty}
u_{1}^{*}(x)u_{2}(x)\,dx,
\end{equation}
so Eq.~\eqref{eq:2.37} establishes the unitary equivalence
\[
\mathscr U:L^{2}(\mathbb R,r(x)dx)\longrightarrow L^{2}(\mathbb R,dx),
\qquad (\mathscr U R)(x)=\sqrt{r(x)}R(x),
\]
between the weighted radial representation and the standard unweighted
Hilbert space. The inverse transformation is $R=u/\sqrt{r(x)}$. The
radial problem therefore takes the one-dimensional Schr\"odinger form}
\begin{equation}
\left[
-\frac{d^{2}}{dx^{2}}
+
V_{\mathrm{eff}}(x)
\right]
u_{\ell}(x)
=
\Lambda_{\ell}(E)u_{\ell}(x),
\label{eq:2.39}
\end{equation}
where the effective potential is (see Figure \ref{fig:3})
\begin{equation}
V_{\mathrm{eff}}(x)
=
\alpha_{B}^{2}\left(x^{2}+b_{0}^{2}\right)
+
\frac{\ell^{2}}{x^{2}+b_{0}^{2}}
+
\frac{2b_{0}^{2}-x^{2}}
{4\left(x^{2}+b_{0}^{2}\right)^{2}}.
\label{eq:2.40}
\end{equation}
{For the remainder of the analysis we use the dimensionless variables
\begin{equation}
\xi=\frac{x}{b_0},\qquad a=\alpha_Bb_0^2,\qquad L=\ell^2,
\label{eq:dimensionless_parameters}
\end{equation}
so that $a>0$ is the confining magnetic parameter and $L$ is the squared azimuthal quantum number.}
The three contributions in Eq.~\eqref{eq:2.40} have distinct origins.
The first term is generated by the quadratic part of the azimuthal magnetic coupling
and provides harmonic confinement for every nonzero magnetic scale
\(\alpha_{B}\). The second term is the regularized centrifugal contribution associated with the
azimuthal quantum number. The third term is purely geometric and originates from the nontrivial radial measure when the weighted radial operator is transformed to the standard unweighted one-dimensional form. Unlike an ordinary centrifugal singularity, the second and third terms remain finite at the Ellis
throat because \(b_{0}>0\). Thus the effective potential at the throat is
\begin{equation}
V_{\mathrm{eff}}(0)
=
\alpha_{B}^{2}b_{0}^{2}
+
\frac{\ell^{2}}{b_{0}^{2}}
+
\frac{1}{2b_{0}^{2}}.
\end{equation}
The throat therefore introduces no singularity into the one-dimensional potential. {For Fig.~\ref{fig:3}, we use the dimensionless variables in Eq.~\eqref{eq:dimensionless_parameters} and the dimensionless potential $\bar V(\xi)\equiv b_0^2V_{\mathrm{eff}}(b_0\xi)$.}

\section{Solvability of the corresponding 1D Schr\"odinger-like equation (\ref{eq:2.39})}
\label{sec:radial_reduction}

{We rewrite the one-dimensional Schrödinger equation \eqref{eq:2.39} after separating the constant magnetic contribution:
\begin{equation}
\begin{split}
&u''_\ell(x)+\left[\widetilde\Lambda_\ell(E)-\widetilde V_{\rm eff}(x)\right]u_\ell(x)=0,\\
&\widetilde V_{\rm eff}(x)=\alpha_B^2x^2+
\frac{\ell^2}{x^2+b_0^2}
+\frac{2b_0^2-x^2}{4(x^2+b_0^2)^2},\\
&\widetilde\Lambda_\ell(E)=\Lambda_\ell(E)-\alpha_B^2b_0^2.
\end{split}
\label{eq:5.1}
\end{equation}
Here \(\Lambda_\ell(E)=E^2-m^2+2\sigma_B\alpha_B\ell\), as defined in Sec.~\ref{sec:charged_KG}. We introduce
\begin{equation}
u_\ell(x)=(x^2+b_0^2)^{1/4}e^{-\alpha_Bx^2/2}H(y),
\qquad y=-\frac{x^2}{b_0^2}.
\label{eq:5.2}
\end{equation}
The resulting equation is
\begin{equation}
{\begin{split}
&(y-1)H''(y)+\left[a y+P-\frac{\beta+1}{y}\right]H'(y)
+\left[(\mu+\nu)-\frac{\mu}{y}\right]H(y)=0,\\
&a=\alpha_Bb_0^2,\quad \beta=\gamma=-\frac12,\quad P=1-a,\quad
(\mu+\nu)=\frac{2a-\widetilde\Lambda_\ell b_0^2}{4},\\
&\mu=\frac{a-\widetilde\Lambda_\ell b_0^2+\ell^2}{4}
\,\,\Rightarrow\,\,
\nu=\frac{a-\ell^2}{4}.
\end{split}}
\label{eq:5.3}
\end{equation}
This is the confluent-Heun equation, whose standard local forms and polynomial-termination properties are discussed in Refs.~\cite{Heun1,Fiziev2010,Levai}. In the standard \(\mathrm{HeunC}\) parametrization its local solution about \(y=0\) may be written as
\begin{equation}
\begin{split}
&H(y)=C_1\,\mathrm{HeunC}(a,\beta,\gamma,\delta,\eta,y)
+C_2\,y^{-\beta}\mathrm{HeunC}(a,-\beta,\gamma,\delta,\eta,y),\\
&P=\beta+\gamma+2-a,\quad\mu=\frac12(a\beta-\beta\gamma+a-\beta-\gamma)-\eta,\\
&\nu=\frac12(a\gamma+\beta\gamma+a+\beta+\gamma)+\delta+\eta.
\end{split}
\label{eq:5.4}
\end{equation}
To provide comprehensive polynomial analysis it is it is convenient to start with Eq.  \eqref{eq:5.3} and use the Heun series 
\begin{equation}
H(y)=\sum_{j=0}^{\infty}A_jy^{j+\chi}.
\label{eq:5.5}
\end{equation}
This would allow us to obtain \(\chi(\chi+\beta)=0\Rightarrow\chi=0 \text{ or } \chi=-\beta\), 
\begin{equation}
A_1=A_0\,
\frac{\chi(\chi+P-1)-\mu}{(\chi+1)(\chi+\beta+1)}
\label{eq:5.6}
\end{equation}
and the three-term recurrence relation
\begin{equation}
\begin{split}
A_{j+2}&(j+\chi+2)(j+\chi+\beta+2)\\
&=A_{j+1}\left[(j+\chi+1)(j+\chi+P)-\mu\right]\\
&\quad+A_j\left[\mu+\nu+a(j+\chi)\right],
\qquad j\geq0.
\end{split}
\label{eq:5.7}
\end{equation}
The Heun power series is to be truncated to a polynomial of degree \(N=p+1\geq1\) to secure finiteness of an admissible solution, i.e., 
\[H(y)=y^\chi\sum_{j=0}^{N}A_jy^j\,;\quad A_N\neq0,\]where \(A_{N+1}=A_{N+2}=\cdots=0\). Two recursion relations play pivotal role in the process. The first is the \((N-1)^{th}\)-order, at \(j=N-1\), recursion relation
\begin{equation}
\begin{split}
A_{N+1}&(N+\chi+1)(N+\chi+\beta+1)\\
&=A_{N}\left[(N+\chi)(N+\chi+P-1)-\mu\right]\\
&\quad+A_{N-1}\left[\mu+\nu+a(N+\chi-1)\right],
\qquad N\geq1.
\end{split}
\label{eq:5.7.1}
\end{equation}
This recursion relation yields \(A_{N+1}=0\) if and only if 
\begin{equation}
A_N=\frac{\left[\mu+\nu+a(N-1+\chi)\right]}{\left[\mu-(N+\chi)(N+\chi+P-1)\right]}A_{N-1}.
\label{eq:5.7.2}
\end{equation}
On the other hand, the second is the  \(N^{th}\)-order recursion relation
\begin{equation}
\begin{split}
A_{N+2}&(N+\chi+2)(N+\chi+\beta+2)\\
&=A_{N+1}\left[(N+\chi+1)(N+\chi+P)-\mu\right]\\
&\quad+A_{N}\left[\mu+\nu+a(N+\chi)\right].
\end{split}
\label{eq:5.7.3}
\end{equation}
It is clear that, \(A_{N+1}=0\), by  Eq.  \eqref{eq:5.7.2}, and \(A_N\neq0\) but 
\begin{equation}
\mu+\nu+a(N+\chi)=0\Rightarrow \delta=-a\left[N+\chi+\frac{1}{2}(\beta+\gamma+2)\right]\,;\,\,N\geq1,
\label{eq:5.8}
\end{equation}
would immediately imply that \(A_{N+2}=0\). One should notice that the standard confluent-Heun parameter \(\delta\) is retrieved \cite{Heun1} for $\chi=0$.  The Heun series truncates to a polynomial of degree \(N\), therefore. Moreover, Eq.  \eqref{eq:5.7.2} simplifies to
\begin{equation}
A_N=\frac{a\,A_{N-1}}{\left[(N+\chi)^2+\nu\right]}.
\label{eq:5.9}
\end{equation}
Equation~\eqref{eq:5.9} is, notably, an additional coefficient-compatibility condition; it is not replaced by Eq.~\eqref{eq:5.8}. Together, these are necessary and sufficient conditions for termination of the Heun series to a polynomial of degree \(N\). This condition in Eq.~\eqref{eq:5.8}   fixes the spectral parameter but does not, by itself, produce a polynomial. The independent coefficient-compatibility condition in Eq.~\eqref{eq:5.9} is also required and selects discrete admissibility points. Generic confined states therefore need not be finite Heun polynomials.}

{That is, for \(N=p+1\), Eq.~\eqref{eq:5.8} and the definitions in Eq.~\eqref{eq:5.3} give
\begin{equation}
\widetilde\Lambda_\ell b_0^2
=a(4p+6+4\chi).
\label{eq:5.11}
\end{equation}
Consequently, the relativistic energy branches are
\begin{equation}
\begin{split}
E(p,\ell,\chi)=\pm\Big[m^2+\alpha_B^2b_0^2+2\alpha_B(2p+3+2\chi-\sigma_B\ell)\Big]^{1/2}.
\end{split}
\label{eq:5.12}
\end{equation}
Thus the two local Frobenius exponents give
\begin{equation}
E(p,\ell)=
\begin{cases}
\displaystyle\pm\sqrt{m^2+\alpha_B^2b_0^2+
2\alpha_B(2p+3-\sigma_B\ell)},&\chi=0,\\[2mm]
\displaystyle\pm\sqrt{m^2+\alpha_B^2b_0^2+
2\alpha_B(2p+4-\sigma_B\ell)},&\chi=\frac12.
\end{cases}
\end{equation}
The second termination condition \eqref{eq:5.9}, after using
Eq.~\eqref{eq:5.8}, yields
\begin{equation}
A_{p+1}=
\frac{aA_p}{(p+\chi+1)^2+\nu},
\qquad
\nu=\frac{a-\ell^2}{4},
\label{eq:5.13}
\end{equation}
provided the displayed denominator is nonzero. In particular,
\begin{equation}
A_{p+1}=
\begin{cases}
\displaystyle\frac{aA_p}{(p+1)^2+\nu},&\chi=0,\\[2mm]
\displaystyle\frac{aA_p}{(p+\frac32)^2+\nu},&\chi=\frac12.
\end{cases}
\end{equation}
Finally, substituting the energy condition into the initial-series
coefficient \eqref{eq:5.6} gives
\begin{equation}
A_1=A_0\,
\frac{4\chi^2+(4p+5)a-\ell^2}
{4(\chi+1)(\chi+\beta+1)}.
\label{eq:5.14}
\end{equation}
For the two exponents this becomes
\begin{equation}
A_1=
\begin{cases}
\displaystyle A_0\,\frac{(4p+5)a-\ell^2}{2},
&\chi=0,\\[2mm]
\displaystyle A_0\,\frac{1+(4p+5)a-\ell^2}{6},
&\chi=\frac12.
\end{cases}
\end{equation}
The compatibility of Eqs.~\eqref{eq:5.13} and \eqref{eq:5.14}
determines the additional parameter restrictions for each polynomial
degree.}

\subsection{Parametric constraints}
{We first consider degree-one confluent-Heun polynomials, corresponding to \(p=0\) and \(N=1\). Equating the two independent expressions for \(A_1/A_0\) in Eqs.~\eqref{eq:5.13} and \eqref{eq:5.14}, with \(\beta=-1/2\), gives the following admissible magnetic parameters. For the even local branch, \(\chi=0\),}
\begin{equation}
a=\frac{3\ell^2-6\pm2\sqrt{\ell^4-4\ell^2+9}}{5},
\qquad a>0,
\label{eq:5.15}
\end{equation}
whereas for the odd local branch, \(\chi=1/2\),
\begin{equation}
a=\frac{3\ell^2-11\pm2\sqrt{\ell^4-4\ell^2+19}}{5},
\qquad a>0.
\label{eq:5.16}
\end{equation}
{Both algebraic branches are retained whenever they give a positive
magnetic parameter. For some integer values of $|\ell|$, both roots are
positive; they correspond to distinct polynomial states and must not be
collapsed into a single branch. The lower positive root gives the
nodeless even degree-one benchmark at $|\ell|=3$ and the lowest odd
degree-one benchmark at $|\ell|=4$ used below. The upper roots have
additional radial nodes at those parameter values. A root with $a=0$ is
excluded because the present confining problem requires $a>0$; roots for which a
recurrence denominator vanishes are likewise excluded from these formulas.}
\subsection{Degree-one Confluent Heun polynomial correspondence}
For \(p=0\), the polynomial factor has degree one:
\begin{equation}
H(y)=y^\chi(A_0+A_1y),\qquad A_0=1.
\label{eq:5.17}
\end{equation}
Substitution into Eq.~\eqref{eq:5.3} gives, successively, the
coefficient conditions
\begin{equation}
\begin{split}
&\mu+\nu=-a(\chi+1),\\
&A_1=\frac{a}{\chi^2+(P+1)\chi+P-\mu},\\
&A_1=\frac{\chi^2+(P-1)\chi-\mu}
{(\chi+1)(\chi+\beta+1)},\\
&\chi(\chi+\beta)=0.
\end{split}
\label{eq:5.17.1}
\end{equation}
{The first and last equations are the spectral and indicial conditions,
respectively; equality of the two expressions for \(A_1\) is the
remaining coefficient-compatibility condition. With \(P=1-a\), this
gives
\begin{equation}
A_1=
\begin{cases}
\displaystyle\frac{4a}{4+a-\ell^2},&\chi=0,\\[2mm]
\displaystyle\frac{4a}{9+a-\ell^2},&\chi=\frac12,
\end{cases}
\label{eq:5.18}
\end{equation}
where \(a\) is fixed by Eq.~\eqref{eq:5.15} or
Eq.~\eqref{eq:5.16}, respectively. The corresponding energies follow
from Eq.~\eqref{eq:5.12} as}
{\begin{equation}
E(0,\ell)=
\begin{cases}
\displaystyle\pm\sqrt{m^2+\alpha_B^2b_0^2+
2\alpha_B(3-\sigma_B\ell)},&\chi=0,\\[2mm]
\displaystyle\pm\sqrt{m^2+\alpha_B^2b_0^2+
2\alpha_B(4-\sigma_B\ell)},&\chi=\frac12.
\end{cases}
\label{eq:5.20}
\end{equation}
Because \(y=-x^2/b_0^2\), the local factor \(y^{1/2}\) is proportional
to \(|x|\) on the real axis and does not, by itself, define a globally
odd function. The global odd mode is obtained by choosing the signed
factor \(x/b_0\) multiplying the corresponding polynomial in
\(y=-x^2/b_0^2\). Thus the \(\chi=0\) and \(\chi=1/2\) branches are
realized globally as even and odd modes, respectively, with the
appropriate signed continuation across the throat.}
\subsection{Degree-two Confluent Heun polynomial correspondence}
For \(p=1\), the confluent-Heun polynomial has degree two. We write it as
\begin{equation}
H(y)=y^\chi(1+A_1y+A_2y^2),\qquad A_0=1.
\label{eq:5.21}
\end{equation}
Using the first termination condition \eqref{eq:5.8}, the recurrence
relations give
\begin{equation}
\begin{split}
A_2&=\frac{4aA_1}
{4\chi^2+16\chi+16+a-\ell^2},\\
A_2&=\frac{A_1(4\chi^2+8\chi+5a+4-\ell^2)-8a}
{4\chi^2+14\chi+12},\\
A_1&=\frac{4\chi^2+9a-\ell^2}
{4\chi^2+6\chi+2}.
\end{split}
\label{eq:5.21.1}
\end{equation}
{The first two expressions for \(A_2\) arise from the recurrence at
\(j=1\) and \(j=0\), respectively, while the last expression is the
initial-coefficient relation. Equating the two expressions for
\(A_2\) imposes the second, independent polynomial-termination
condition. The local exponents are again \(\chi=0\) and \(\chi=1/2\);
the latter must be continued with a signed factor \(x/b_0\) to
represent a globally odd mode on the complete wormhole.}

{For degree two, the compatibility condition is obtained by equating the two expressions for \(A_2\) in Eq.~\eqref{eq:5.21.1}, after substituting the corresponding expression for \(A_1\). For even parity, \(\chi=0\), this gives
\begin{equation}
L^3-20L^2-15L^2a+64L
+200La+59La^2
-320a-308a^2-45a^3=0.
\label{eq:5.22}
\end{equation}
For odd parity, \(\chi=1/2\), the condition is
\begin{equation}
L^3-35L^2-15L^2a+259L
+318La+59La^2
-225-879a-443a^2-45a^3=0.
\label{eq:5.23}
\end{equation}
These are cubic equations in \(a\), with $L=\ell^2$; only positive real roots correspond to a nonzero magnetic coupling. Representative admissible roots are: }
\begin{equation}
\begin{split}
\chi=0:\quad&
\ell=1:\ a=0.230896674674,\\
&\ell=2:\ a=1.644040370643,\\
&\ell=3:\ a=5.766291075008,
\end{split}
\label{eq:5.24}
\end{equation}
and
\begin{equation}
\begin{split}
\chi=\frac12:\quad&
\ell=2:\ a=1.400000000000,\\
&\ell=3:\ a=5.223094732892.
\end{split}
\label{eq:5.25}
\end{equation}
For the odd branch, the cases \(\ell=0,1\), have no positive real
solution. For the even branch, \(\ell=0\) has no positive solution.
The examples above are representative admissibility points, not a
complete enumeration of all roots for every integer \(\ell\). The
corresponding relativistic energies are
\begin{equation}
E(1,\ell)=
\begin{cases}
\displaystyle\pm\sqrt{m^2+\alpha_B^2b_0^2+
2\alpha_B(5-\sigma_B\ell)},&\chi=0,\\[2mm]
\displaystyle\pm\sqrt{m^2+\alpha_B^2b_0^2+
2\alpha_B(6-\sigma_B\ell)},&\chi=\frac12.
\end{cases}
\label{eq:5.26}
\end{equation}
Here \(a\), defined in Eq.~\eqref{eq:dimensionless_parameters}, is selected from the positive roots of
Eqs.~\eqref{eq:5.22} and \eqref{eq:5.23}.

\subsection{Fourth-order finite-difference validation and higher radial excitations}
\label{sec:fd_validation}

{The analytic degree-one modes provide stringent spectral benchmarks, but they do not determine the generic confined spectrum. We therefore solve the full-line eigenvalue problem independently by a symmetric finite-difference discretization and retain several excited levels in each parity sector. Using the dimensionless variables in Eq.~\eqref{eq:dimensionless_parameters}, we further define the dimensionless spectral parameter
\begin{equation}
\varepsilon=b_0^2\Lambda_\ell(E).
\label{eq:fd_dimensionless_variables}
\end{equation}
Then Eqs.~\eqref{eq:2.39} and \eqref{eq:2.40} become
\begin{equation}
\left[-\frac{d^2}{d\xi^2}+W_{aL}(\xi)\right]u(\xi)
=\varepsilon u(\xi),
\label{eq:fd_dimensionless_schrodinger}
\end{equation}
where
\begin{equation}
W_{aL}(\xi)=a^2(1+\xi^2)+\frac{L}{1+\xi^2}
+\frac{2-\xi^2}{4(1+\xi^2)^2}.
\label{eq:fd_dimensionless_potential}
\end{equation}
Here $W_{aL}(\xi)\equiv\bar V(\xi)$ is the dimensionless effective potential introduced in Fig.~\ref{fig:3}; the notation $W_{aL}$ makes its dependence on $a$ and $L$ explicit. We denote the ordered discrete eigenvalues of this full-line operator by $\varepsilon_0<\varepsilon_1<\cdots$, with corresponding eigenfunctions $u_n(\xi)$.}

\begin{table*}[t]
\centering
 \caption{\fontsize{7.6}{8.6}\selectfont First eight dimensionless full-line eigenvalues $\varepsilon_n$ obtained by fourth-order finite differences and Richardson extrapolation. The parity and node count refer to the global eigenfunction on $\mathbb R$. The refinement correction is $\Delta_n^{\mathrm R}=|\varepsilon_n^{\mathrm R}-\varepsilon_n(h=0.005)|$; it is a convergence diagnostic, not a rigorous error bound. The two benchmark groups correspond to the even and odd degree-one analytic parameter points, respectively; the listed levels in each group include both global parities.}
\label{tab:fd_spectra}
\setlength{\tabcolsep}{3pt}
\scriptsize
\begin{tabular}{ccclrrl}
\hline\hline
Benchmark & $n$ & Parity & Nodes & $\varepsilon_n^{\mathrm R}$ & $\Delta_n^{\mathrm R}$ & Status\\
\hline
Even degree-one benchmark, $|\ell|=3$ & 0 & even & 0 & $9.152817244713$ & $6.78\times10^{-11}$ & analytic $^*$\\
& 1 & odd & 1 & $9.652594424548$ & $8.77\times10^{-11}$ & numerical\\
& 2 & even & 2 & $12.292878099693$ & $2.99\times10^{-10}$ & numerical\\
& 3 & odd & 3 & $13.903885537506$ & $8.06\times10^{-10}$ & numerical\\
& 4 & even & 4 & $16.243115178554$ & $1.81\times10^{-9}$ & numerical\\
& 5 & odd & 5 & $18.401620449341$ & $3.63\times10^{-9}$ & numerical\\
& 6 & even & 6 & $20.747498549066$ & $6.37\times10^{-9}$ & numerical\\
& 7 & odd & 7 & $23.072681204153$ & $1.03\times10^{-8}$ & numerical\\
\hline
Odd degree-one benchmark, $|\ell|=4$ & 0 & even & 0 & $14.967865805794$ & $1.61\times10^{-10}$ & numerical\\
& 1 & odd & 1 & $15.244347897422$ & $1.84\times10^{-10}$ & analytic $^*$\\
& 2 & even & 2 & $18.941792398802$ & $6.53\times10^{-10}$ & numerical\\
& 3 & odd & 3 & $20.366025250504$ & $1.39\times10^{-9}$ & numerical\\
& 4 & even & 4 & $23.319285488007$ & $3.00\times10^{-9}$ & numerical\\
& 5 & odd & 5 & $25.761632862651$ & $6.13\times10^{-9}$ & numerical\\
& 6 & even & 6 & $28.627252967300$ & $1.08\times10^{-8}$ & numerical\\
& 7 & odd & 7 & $31.425587222204$ & $1.79\times10^{-8}$ & numerical\\
\hline\hline
\end{tabular}
\end{table*}

\begin{table*}[t]
\centering
\caption{\fontsize{7.6}{8.6}\selectfont Representative lowest full-line eigenvalues of the dimensionless operator in Eq.~\eqref{eq:fd_dimensionless_schrodinger}, obtained with the same fourth-order discretization at $X=8$ and $h=0.005$. The entries are finite-grid eigenvalues; the last column gives the corresponding exact degree-zero analytic checks where available. The numerical value and the continuum eigenvalue are kept distinct.}
\label{tab:fd_representative_spectra}
\setlength{\tabcolsep}{5pt}
\scriptsize
\begin{tabular}{ccclrrrr}
\hline\hline
$a$ & $L$ & sector & $n=0$ & $n=1$ & $n=2$ & $n=3$ & analytic check\\
\hline
$0.5$ & $0$ & generic & $0.9495230336$ & $1.7788273770$ & $2.8185632178$ & $3.7786325859$ & --\\
$1$ & $1$ & even degree zero & $3.0000000000$ & $4.5495162566$ & $6.5029473686$ & $8.4065044763$ & $a=L,\ \varepsilon_0=3$\\
$1$ & $4$ & generic & $4.99999999999$ & $5.8497004347$ & $7.8052727784$ & $9.4441371932$ & --\\
$3$ & $4$ & odd degree zero & $15.8688721609$ & $20.9999999983$ & $26.5923546912$ & $32.3063238461$ & $a=L-1,\ \varepsilon_1=21$\\
\hline\hline
\end{tabular}
\end{table*}

\begin{table*}[t]
\centering
\caption{\fontsize{7.6}{8.6}\selectfont Degree-two even Heun-polynomial checks at integer azimuthal numbers. The algebraic parameter is selected by the even degree-two admissibility condition in Eq.~\eqref{eq:5.22}, with the representative positive roots listed in Eq.~\eqref{eq:5.24}. For each listed root, the polynomial has two positive roots in $\xi^2$ and hence four full-line nodes; $\varepsilon_4^{\rm exact}=a^2+10a$. The numerical column is the fourth-order Richardson estimate using $X=20$ for $|\ell|=1$ and $X=8$ for $|\ell|=2,3$, using the two finest meshes $h=0.01$ and $h=0.005$.}
\label{tab:degree_two_checks}
\setlength{\tabcolsep}{6pt}
\scriptsize
\begin{tabular}{crrrr}
\hline\hline
$|\ell|$ & $a$ & $\varepsilon_4^{\rm exact}$ & $\varepsilon_4^{\rm FD,R}$ & absolute difference\\
\hline
1 & $0.2308966746735715$ & $2.36228002111103$ & $2.36228002112227$ & $1.1\times10^{-11}$\\
2 & $1.6440403706431146$ & $19.14327244673549$ & $19.14327244673545$ & $5.0\times10^{-14}$\\
3 & $5.766291075007751$ & $90.91302351179158$ & $90.91302351173320$ & $5.8\times10^{-11}$\\
\hline\hline
\end{tabular}
\end{table*}

{The dimensionless potential can develop a symmetric double well when the centrifugal contribution is sufficiently strong. Set $t=\xi^2$ and write
\begin{equation}
W_{aL}(t)=a^2(1+t)+\frac{L}{1+t}+\frac{2-t}{4(1+t)^2}.
\label{eq:double_well_Wt}
\end{equation}
Its throat slope is
\begin{equation}
\left.\frac{dW_{aL}}{dt}\right|_{t=0}=a^2-L-\frac54.
\label{eq:double_well_throat_slope}
\end{equation}
Thus $L>a^2-5/4$ makes the throat a local maximum. The off-throat minima are determined by the positive roots of
\begin{equation}
4a^2(1+t_\star)^3-4L(1+t_\star)+t_\star-5=0,
\qquad t_\star>0,
\label{eq:double_well_minima}
\end{equation}
and the barrier height is $\Delta W_{\rm b}=W_{aL}(0)-W_{aL}(t_\star)$. When the two minima are well separated, the lowest even and odd eigenstates are approximately the symmetric and antisymmetric combinations of states localized near the two minima. Their splitting $\Delta\varepsilon=\varepsilon_1-\varepsilon_0$ has the leading semiclassical exponential dependence determined by the forbidden-region action
\begin{equation}
\mathcal S(\varepsilon_0)=
\int_{-\xi_{\rm in}}^{\xi_{\rm in}}
\sqrt{W_{aL}(\xi)-\varepsilon_0}\,d\xi,
\qquad
\Delta\varepsilon\ \propto\ e^{-\mathcal S},
\label{eq:double_well_action}
\end{equation}
where $\pm\xi_{\rm in}$ are the inner turning points satisfying $W_{aL}(\xi_{\rm in})=\varepsilon_0$. The action determines the leading exponential dependence of the level splitting; its prefactor is not evaluated here.For every $a>0$, this potential is smooth, real, even, and confining, with
$W_{aL}(\xi)\sim a^2\xi^2$ as $|\xi|\to\infty$. The numerical problem is consequently posed on the complete line, not on a single side of the throat. Reflection symmetry implies that the eigenfunctions alternate in parity: the state with $n$ interior nodes has parity $(-1)^n$. In particular, the even ground state has no node, while the lowest odd state has its unique node at $\xi=0$. The throat conditions $u'(0)=0$ and $u(0)=0$ characterize the even and odd eigenfunctions, respectively; they are not additional outer boundary conditions. The node ordering follows the one-dimensional Sturm oscillation theorem for confining Schr\"odinger operators \cite{Simon2005}.}

{\subsection{Physical interpretation: magnetic response, two-end localization, and spectral hybridization}}
{The radial spectrum receives contributions from the magnetic coupling, the azimuthal quantum number, and the geometry. First, the applied field enters the radial operator through the non-negative scale $\alpha_B=|q\mathcal B_0|/2$. Its contribution $\alpha_B^2r^2(x)$ grows without bound at either asymptotic end and confines the scalar mode. This is confinement generated by azimuthal minimal coupling on the curved surface, not a gravitational potential generated by the wormhole. The geometry modifies the finite-radius spectrum through the regularized angular term and the curvature-induced term associated with the radial measure.}

{Second, the two asymptotic ends are not separate radial systems. The proper coordinate $x$ covers both ends in one connected manifold, and the effective potential is invariant under $x\mapsto-x$. Every nondegenerate bound state can therefore be chosen with definite global parity. An even mode has equal-sign radial amplitudes on the two sides, whereas an odd mode has opposite signs and vanishes at the throat. The positive localization density is even for either parity; a unit-normalized mode consequently has one half of its integrated spatial weight in each asymptotic end. This equal-weight statement describes one global state on the connected wormhole, not two independent or degenerate states.}

{Third, when the effective potential has two off-throat minima separated by a throat maximum, the lowest even and odd modes can, in the well-separated regime, be viewed as symmetric and antisymmetric combinations of configurations localized predominantly in the two mouths. Their energy separation measures inter-mouth hybridization. The forbidden-region action gives the leading semiclassical exponential dependence of this splitting, whereas its prefactor and any time-dependent transfer probability require a separate calculation and are not inferred here. Outside the well-separated regime, parity remains exact, but a localized-left/localized-right interpretation need not be quantitatively useful.}

{The magnetic coupling also separates radial-profile dependence from the relativistic energy shift. The one-dimensional operator depends on $\alpha_B$ and $L=\ell^2$, so the radial profiles are insensitive to the sign of $q\mathcal B_0$ at fixed magnitude and fixed $|\ell|$. The sign-sensitive orbital cross term instead enters through $\Lambda_\ell(E)=E^2-m^2+2\sigma_B\alpha_B\ell$, shifting the relativistic energy branches according to the relative sign of $\sigma_B$ and $\ell$. Thus the field magnitude controls confinement, while the relative orientation of charge-field coupling and azimuthal angular momentum controls the energy offset. Only branches with a nonnegative radicand correspond to real frequencies (see Eq.~\eqref{eq:fd_relativistic_energies}).}

{Finally, the throat holonomy is the gauge-invariant phase accumulated around the azimuthal circle. It characterizes the global connection and can remain nontrivial when the local oriented magnetic scalar vanishes at the throat. The holonomy and the local magnetic scalar therefore encode complementary global and local properties of the prescribed electromagnetic background.}

{For the discretization, we truncate the line to $-X\leq\xi\leq X$ and impose Dirichlet conditions only at the remote endpoints, $u(-X)=u(X)=0$. On a uniform mesh $\xi_j=-X+jh$, the second derivative is represented by the fourth-order, five-point central formula
\begin{equation}
u''(\xi_j)=
\frac{-u_{j+2}+16u_{j+1}-30u_j+16u_{j-1}-u_{j-2}}
{12h^2}+O(h^4).
\label{eq:fd_five_point_stencil}
\end{equation}
The coefficients are the standard centered fourth-order finite-difference weights \textcolor{revUnified}{\cite{Fornberg1988}}.
The resulting real symmetric pentadiagonal matrix is diagonalized for its lowest eigenvalues, following established finite-difference treatments of one-dimensional Schrödinger eigenvalue problems \textcolor{revUnified}{\cite{Simos1998}}. At the first and last interior grid points, the five-point stencil is closed by odd reflection of the Dirichlet data, $u_{-1}=-u_1$ and $u_{N+1}=-u_{N-1}$ when the endpoints are indexed by $j=0,N$. Thus the first and last diagonal entries are $29/(12h^2)+W_{aL}(\xi_j)$, while interior diagonal entries are $30/(12h^2)+W_{aL}(\xi_j)$; the off-diagonal entries retain the five-point coefficients. This specifies the boundary rows of the symmetric pentadiagonal matrix. Because the bound-state amplitudes are exponentially suppressed at $|\xi|=X$, the endpoint closure has negligible influence on the reported levels, as verified by the independent interval variation below. We use $X=8$ and $h=0.005$ for the reported spectra, with $h=0.01$ as the refinement level. The interval is independently varied over $X=7,8,9$. At $h=0.005$, changing the interval from $[-7,7]$ to $[-9,9]$ changes the first eight eigenvalues by less than $4.0\times10^{-11}$ at either benchmark point. This variation is at the level of the floating-point diagonalization and shows no systematic endpoint drift; the reported refinement diagnostic is controlled by mesh discretization and eigensolver precision; endpoint effects are below this level. The fourth-order convergence is checked by halving $h$; the leading error is proportional to $h^4$, and we use the Richardson extrapolation estimate \textcolor{revUnified}{\cite{Bolton1956}}
\begin{equation}
\varepsilon_n^{\mathrm R}=
\frac{16\varepsilon_n(h/2)-\varepsilon_n(h)}{15},
\qquad h=0.01,
\label{eq:fd_richardson}
\end{equation}
to remove the leading discretization error. The values in Table~\ref{tab:fd_spectra} are these extrapolated estimates.}

{Two independent analytic benchmark points are used. For the even degree-one mode, $|\ell|=3$ and
\begin{equation}
\begin{aligned}
a&=a_{\mathrm e}^{(-)}(9)=\frac{21-2\sqrt{54}}{5},\\
\varepsilon_{\mathrm e}^{\mathrm{exact}}&=a^2+6a
=9.15281724470668\ldots .
\end{aligned}
\label{eq:fd_even_benchmark}
\end{equation}
For the odd degree-one mode, $|\ell|=4$ and
\begin{equation}
\begin{aligned}
a&=a_{\mathrm o}^{(-)}(16)=\frac{37-2\sqrt{211}}{5},\\
\varepsilon_{\mathrm o}^{\mathrm{exact}}&=a^2+8a
=15.24434789743438\ldots .
\end{aligned}
\label{eq:fd_odd_benchmark}
\end{equation}
The second benchmark is the first odd-parity level, not the ground state of the full operator. The full-line ground state at that same parameter point is even and lies below it. This ordering is a direct numerical check of the oscillation theorem and prevents the odd polynomial solution from being misidentified as a global ground state.}

{A coarser $h=0.02$ calculation was also performed. Across the first eight levels at both benchmark points, the ratios $[\varepsilon_n(h)-\varepsilon_n(h/2)]/[\varepsilon_n(h/2)-\varepsilon_n(h/4)]$, evaluated at $h=0.02$, range from $15.92$ to $16.07$, consistent with the fourth-order limit of $16$ and the expected finite-$h$ corrections. The small departure from the ideal value $16$ reflects higher-order and floating-point/eigensolver effects. The Richardson-extrapolated values reproduce the exact even and odd benchmark eigenvalues within $1.3\times10^{-11}$; this residual is comparable to the floating-point sensitivity of diagonalizing matrices whose kinetic entries scale as $h^{-2}$. The raw $h=0.005$ values already agree with the exact benchmarks to better than $2\times10^{-10}$. These comparisons validate both the differential operator and the parity assignment independently of the Heun-polynomial construction.}

{Table~\ref{tab:fd_spectra} gives the first eight full-line levels at the two degree-one benchmark points, while Table~\ref{tab:fd_representative_spectra} extends the comparison to additional magnetic and angular parameters, including both degree-zero analytic sectors; the additional degree-two checks at $|\ell|=1,2,3$ are given in Table~\ref{tab:degree_two_checks}. Here $n$ is the number of interior nodes, so the parity is even for even $n$ and odd for odd $n$. {Figure~\ref{fig:4}(c) displays the even degree-one benchmark potential and its first eight full-line levels, while Fig.~\ref{fig:4}(d) shows the corresponding odd degree-one benchmark; the plotted level values are those listed in Table~\ref{tab:fd_spectra}, not additional eigenvalue calculations.} The tabulated eigenvalues are the Richardson estimates in Eq.~\eqref{eq:fd_richardson}. At the even benchmark, the exact degree-one solution is the $n=0$ ground state. At the odd benchmark, the exact degree-one solution is the $n=1$ level; the $n=0$ even state is lower, as required for a one-dimensional confining operator. }

{The numerical spectrum follows the parity alternation required by the one-dimensional oscillation theorem. Each degree-one analytic benchmark is compared with the numerical spectrum at the same admissible pair $(a,L)$.}

{The parity of every computed eigenvector was checked using its reflection overlap. This quantity is the discrete-grid representation of the normalized expectation value of the spatial parity operator, which acts on a wavefunction by reflection, $(\hat{\mathcal P}u)(x)=u(-x)$ \cite{Zettili2009}:
\begin{equation}
\mathcal P_n=
\frac{\sum_j u_{n,j}u_{n,-j}}{\sum_j|u_{n,j}|^2},
\label{eq:fd_parity_overlap}
\end{equation}
For the first eight levels at both parameter points, $\mathcal P_n$ equals $+1$ for even $n$ and $-1$ for odd $n$, to within $2.3\times10^{-16}$. The node count was independently obtained from sign changes after excluding tail values smaller than $10^{-7}$ of the maximum amplitude; it gives exactly $n$ for all sixteen reported states. These checks verify the global parity and Sturm ordering, including the throat node of every odd state.}

{Finally, the numerical eigenvalues translate directly into the relativistic frequency branches through the spectral definition in Eq.~\eqref{eq:2.33a}:
\begin{equation}
\begin{aligned}
(E_nb_0)^2&=(mb_0)^2+\varepsilon_n-2\sigma_Ba\ell,\\
E_n^{(\pm)}&=\pm b_0^{-1}
\sqrt{(mb_0)^2+\varepsilon_n-2\sigma_Ba\ell}.
\end{aligned}
\label{eq:fd_relativistic_energies}
\end{equation}
provided the radicand is nonnegative. Thus the tables report the geometry- and magnetic-field-dependent radial spectrum before the charge-sign and frequency-branch choice. The numerical calculation verifies the degree-one analytic levels and determines higher confined excitations, whose eigenfunctions are generally non-polynomial confluent-Heun solutions.}

{The global spectral ordering is fixed by the one-dimensional oscillation theorem. At the lower positive even degree-one root in Eq.~\eqref{eq:5.15}, the polynomial factor has no real zero and the state is the full-line ground state. At the lower positive odd degree-one root in Eq.~\eqref{eq:5.16}, the signed factor contributes the single throat node and gives the lowest odd-parity state. These identifications are confirmed by the finite-difference benchmarks in Eqs.~\eqref{eq:fd_even_benchmark} and \eqref{eq:fd_odd_benchmark}. Away from the polynomial-admissibility loci, confinement and discreteness persist, but the eigenfunctions are generally non-polynomial solutions of the confluent-Heun spectral problem.}

\subsection{Lowest analytic modes, exact normalization, and radial probability densities}
\label{sec:normalized_lowest_modes}

\begin{figure*}[t]
\centering
\includegraphics[width=0.86\textwidth]{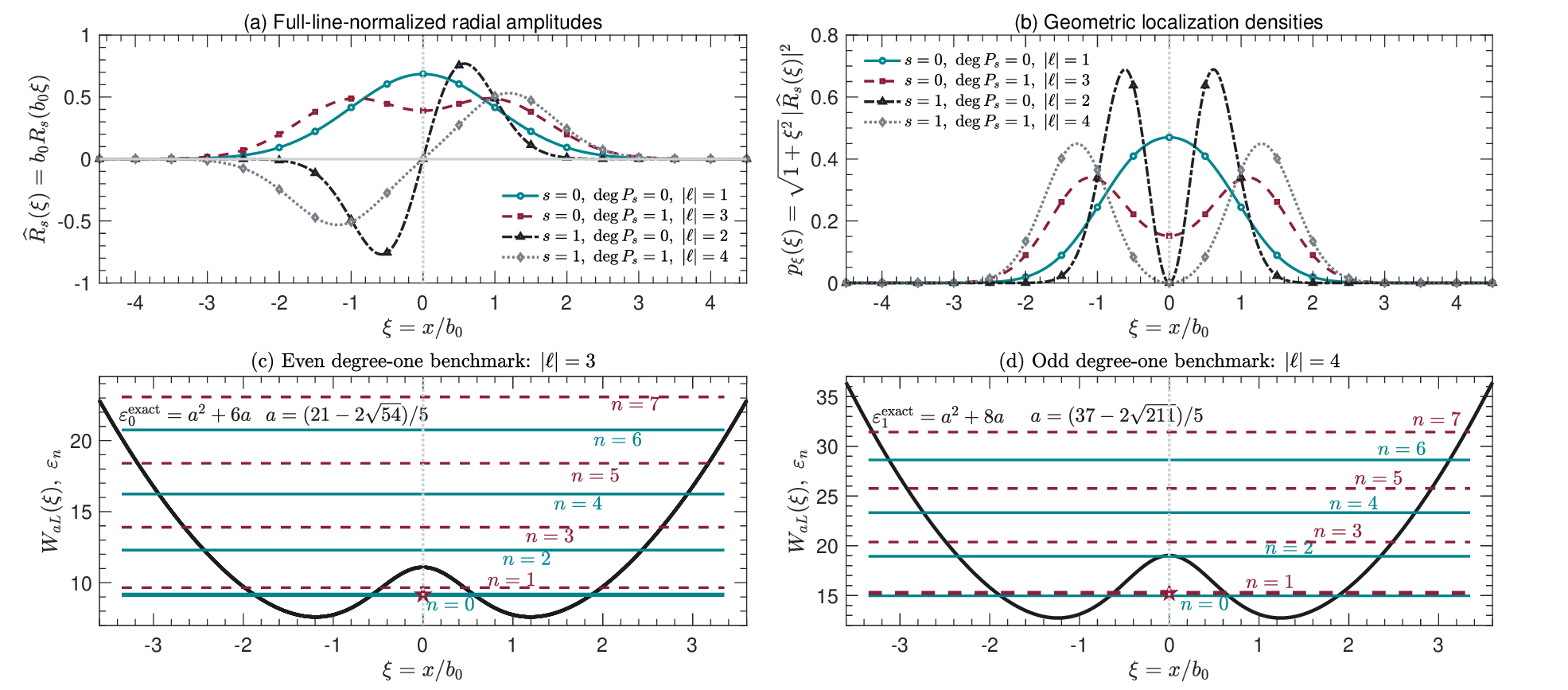}
\caption{\fontsize{7.6}{8.6}\selectfont
Four-panel representation of the normalized polynomial modes and the corresponding full-line spectral benchmarks on the Ellis wormhole. The dimensionless coordinate and magnetic parameter are $\xi=x/b_0$ and $a=\alpha_Bb_0^2$, respectively, and $L=\ell^2$. (a) The full-line-normalized radial amplitudes $\widehat R_s(\xi)=b_0R_s(b_0\xi)$ (see Eqs.~\eqref{eq:norm_general_low_modes}--\eqref{eq:norm_constants_low_modes}): the degree-zero even mode uses $(s,\ell,a)=(0,1,1)$; the degree-zero odd mode uses $(s,\ell,a)=(1,2,3)$; the degree-one even mode uses $|\ell|=3$, $L=9$, and $a=(21-2\sqrt{54})/5=1.260612308660$; and the degree-one odd mode uses $|\ell|=4$, $L=16$, and $a=(37-2\sqrt{211})/5=1.589664381466$. The degree-one polynomial coefficients are $c_{\rm e}=4a/(4+a-L)=-1.348469228350$ and $c_{\rm o}=4a/(9+a-L)=-1.175279682111$, respectively (see Eq.~\eqref{eq:norm_degreeone_coefficients}). The odd profiles are continued with the signed factor $\xi$, ensuring global odd parity. (b) The corresponding positive localization densities $p_\xi(\xi)=\sqrt{1+\xi^2}\,|\widehat R_s(\xi)|^2$, normalized with the Ellis measure (see Eq.~\eqref{eq:norm_probability_xi}). (c) The dimensionless potential $W_{aL}(\xi)$ (see Eq.~\eqref{eq:fd_dimensionless_potential}) and the first eight full-line eigenvalues for the even degree-one benchmark, $|\ell|=3$, $L=9$, and $a=(21-2\sqrt{54})/5$. (d) The corresponding potential and first eight eigenvalues for the odd degree-one benchmark, $|\ell|=4$, $L=16$, and $a=(37-2\sqrt{211})/5$. In (c) and (d), the horizontal lines are the Richardson-extrapolated eigenvalues $\varepsilon_n^{\rm R}$ in Table~\ref{tab:fd_spectra}, with $n=0,\ldots,7$; solid turquoise and dashed burgundy distinguish even and odd global parity, respectively. The burgundy star marks the exact polynomial benchmark, $\varepsilon_{\rm e}^{\rm exact}=a^2+6a$ in (c) and $\varepsilon_{\rm o}^{\rm exact}=a^2+8a$ in (d), as given in Eqs.~\eqref{eq:fd_even_benchmark} and \eqref{eq:fd_odd_benchmark}. The plotted amplitudes and densities use the dimensionless plotting convention $b_0=1$; this is a choice of units for the display and does not fix the physical throat radius. All panels use the full radial domain $\xi\in\mathbb R$; the finite horizontal plotting windows are display ranges only.}
\label{fig:4}
\end{figure*}

{The magnetic confinement problem admits a useful set of closed-form states at the
polynomial-admissibility points identified above. Their normalization is defined by the geometric radial measure. We retain $\xi$, $a$, and $L$ from Eq.~\eqref{eq:dimensionless_parameters}; the circumferential radius is the geometric function $r(x)=\sqrt{x^2+b_0^2}$ already defined in Eq.~\eqref{eq:2.6}.
The inverse unitary transformation is $R_{\ell}=u_{\ell}/\sqrt{r}$
(see Eq.~\eqref{eq:2.37}). Substituting the Heun ansatz
$u_{\ell}=r^{1/2}e^{-\alpha_Bx^2/2}H(-x^2/b_0^2)$,
with $r^{1/2}=(x^2+b_0^2)^{1/4}$, shows explicitly that the geometrical
prefactor cancels when one returns to the physical radial
Klein--Gordon amplitude. Thus
\begin{equation}
R_\ell(x)=\mathcal N\,e^{-a\xi^2/2}H(-\xi^2).
\label{eq:norm_radial_from_heun}
\end{equation}
For a globally defined odd mode, the local Frobenius factor
$y^{1/2}$ is continued as the signed factor $\xi$, not $|\xi|$.
It is therefore convenient to write the lowest even and odd polynomial
families as
\begin{equation}
R_s(x)=\mathcal N_s\,\xi^s e^{-a\xi^2/2}
P_s(\xi^2),\qquad s=0,1,
\label{eq:norm_general_low_modes}
\end{equation}
where $s=0$ and $s=1$ denote global even and odd parity, respectively,
and $P_s$ is a real polynomial. This representation is regular at the
throat and decays at both asymptotic ends for every $a>0$.

For degree zero, $H(y)=y^\chi$ and termination is imposed directly at
$N=0$. The two local exponents give the even and odd admissibility
conditions
\begin{equation}
\begin{aligned}
&s=0:\qquad a=L,\qquad
b_0^2\widetilde\Lambda_\ell=2a,\\
&s=1:\qquad a=L-1>0,\qquad
b_0^2\widetilde\Lambda_\ell=4a.
\end{aligned}
\label{eq:norm_degreezero_conditions}
\end{equation}
Thus the even degree-zero state is nodeless and is the full-line ground
state at its admissible parameter point, whereas the odd degree-zero
state has its required single node at the throat and is the lowest
odd-parity state at its own admissible point. Using
$\Lambda_\ell=\widetilde\Lambda_\ell+\alpha_B^2b_0^2$ and
$\Lambda_\ell=E^2-m^2+2\sigma_B\alpha_B\ell$, their energy branches are
\begin{equation}
\begin{aligned}
E_{0,\mathrm e}^{(\pm)}
&=\pm\sqrt{m^2+\alpha_B^2b_0^2
+2\alpha_B(1-\sigma_B\ell)},\\
E_{0,\mathrm o}^{(\pm)}
&=\pm\sqrt{m^2+\alpha_B^2b_0^2
+2\alpha_B(2-\sigma_B\ell)}.
\end{aligned}
\label{eq:norm_degreezero_energies}
\end{equation}
Only parameter choices for which $a>0$ and the displayed energy radicand
is nonnegative describe magnetically confined modes with real frequency.

The next polynomial family has degree one. In the notation of
Eqs.~\eqref{eq:5.17}--\eqref{eq:5.20}, its global even and odd radial
profiles are
\begin{equation}
\begin{aligned}
R_{1,\mathrm e}(x)
&=\mathcal N_{1,\mathrm e}e^{-a\xi^2/2}
\left(1-c_{\mathrm e}\xi^2\right),\\
R_{1,\mathrm o}(x)
&=\mathcal N_{1,\mathrm o}\xi e^{-a\xi^2/2}
\left(1-c_{\mathrm o}\xi^2\right),
\end{aligned}
\label{eq:norm_degreeone_profiles}
\end{equation}
where
\begin{equation}
c_{\mathrm e}=\frac{4a}{4+a-L},\qquad
c_{\mathrm o}=\frac{4a}{9+a-L}.
\label{eq:norm_degreeone_coefficients}
\end{equation}
The corresponding values of $a$ must satisfy the appropriate even or odd
degree-one compatibility condition in Sec.~\ref{sec:radial_reduction};
the coefficients in Eq.~\eqref{eq:norm_degreeone_coefficients} are not
independent parameters. Their relativistic energies are those in
Eq.~\eqref{eq:5.20}. The number of nodes is determined by the complete
polynomial factor: polynomial degree alone does not determine the
ordering of a state in the full-line spectrum.

For exact normalization, define the positive moments
\begin{equation}
\begin{aligned}
\mathcal J_k(a)
&=\int_{-\infty}^{\infty}
\xi^{2k}\sqrt{1+\xi^2}\,e^{-a\xi^2}\,d\xi\\
&=\int_{0}^{\infty}t^{k-\frac12}(1+t)^{\frac12}e^{-at}\,dt\\
&=\Gamma\!\left(k+\frac12\right)
U\!\left(k+\frac12,k+2,a\right),
\qquad k=0,1,\ldots,
\end{aligned}
\label{eq:norm_exact_moments}
\end{equation}
where $U(\mathsf A,\mathsf B,z)$ is Tricomi's confluent hypergeometric
function. The second equality follows from $t=\xi^2$; the last is the
standard integral representation of $U$ in \textcolor{revUnified}{\cite{NISTDLMF}},
valid here because $a>0$. The natural radial normalization,
\begin{equation}
\int_{-\infty}^{\infty}r(x)|R_\ell(x)|^2\,dx=1,
\label{eq:norm_radial_condition}
\end{equation}
then gives
\begin{equation}
\begin{aligned}
\mathcal N_{0,\mathrm e}
&=\frac{1}{b_0\sqrt{\mathcal J_0(a)}},
&
\mathcal N_{0,\mathrm o}
&=\frac{1}{b_0\sqrt{\mathcal J_1(a)}},\\[1mm]
\mathcal N_{1,\mathrm e}
&=\frac{1}{b_0
\sqrt{\mathcal J_0-2c_{\mathrm e}\mathcal J_1
+c_{\mathrm e}^{\,2}\mathcal J_2}},\\
\mathcal N_{1,\mathrm o}
&=\frac{1}{b_0
\sqrt{\mathcal J_1-2c_{\mathrm o}\mathcal J_2
+c_{\mathrm o}^{\,2}\mathcal J_3}}.
\end{aligned}
\label{eq:norm_constants_low_modes}
\end{equation}
In the last two lines all moments are evaluated at the same $a$ as the
corresponding mode. These expressions retain the exact Ellis measure
$r(x)\,dx$; replacing it by $dx$ would give different normalization
constants and would not normalize the original radial field.

The associated radial localization density with respect to the proper
radial coordinate is
\begin{equation}
p_x(x)=r(x)|R_\ell(x)|^2=|u_\ell(x)|^2,
\qquad
\int_{-\infty}^{\infty}p_x(x)\,dx=1.
\label{eq:norm_probability_x}
\end{equation}
Equivalently, the dimensionless density with respect to $d\xi$ is
\begin{equation}
p_\xi(\xi)=b_0p_x(b_0\xi)
=\frac{\sqrt{1+\xi^2}\,\xi^{2s}
e^{-a\xi^2}P_s(\xi^2)^2}
{\displaystyle\int_{-\infty}^{\infty}
\sqrt{1+\zeta^2}\,\zeta^{2s}
e^{-a\zeta^2}P_s(\zeta^2)^2\,d\zeta}.
\label{eq:norm_probability_xi}
\end{equation}
In particular, the four lowest polynomial profiles above are obtained
by taking $(s,P_s)=(0,1)$, $(1,1)$, $(0,1-c_{\mathrm e}\xi^2)$,
and $(1,1-c_{\mathrm o}\xi^2)$, respectively. {Their normalized amplitudes and the associated geometric localization densities are shown in Fig.~\ref{fig:4}(a) and Fig.~\ref{fig:4}(b), respectively.} Each density is even,
irrespective of the parity of the underlying amplitude. Consequently,
the normalized probability contained in either asymptotic end is
\begin{equation}
\int_{-\infty}^{0}p_\xi(\xi)\,d\xi
=\int_{0}^{\infty}p_\xi(\xi)\,d\xi
=\frac12.
\label{eq:norm_two_end_probability}
\end{equation}
For an odd mode the density vanishes at the throat, while an even
nodeless mode has its maximum amplitude at or near the throat depending
on the geometric measure and the parameters. The density in Eq.~\eqref{eq:norm_probability_xi} is the positive spatial
localization density associated with the self-adjoint radial problem.
The conserved Klein--Gordon charge density is a distinct quantity and
is generally sign-indefinite.

For completeness, the unit spatial norm in Eq.~\eqref{eq:norm_radial_condition}
can be converted to the standard Klein--Gordon inner-product
normalization. With the angular factor normalized as
$e^{i\ell\phi}/\sqrt{2\pi}$ and with $A_t=0$, a separated mode
$\Psi_{E\ell}=e^{-iEt+i\ell\phi}R_\ell(x)/\sqrt{2\pi}$ has
\begin{equation}
(\Psi_{E\ell},\Psi_{E\ell})_{\rm KG}
=2E\int_{-\infty}^{\infty}r(x)|R_\ell(x)|^2\,dx
=2E.
\label{eq:norm_KG_inner_product}
\end{equation}
Thus a positive-frequency mode with unit Klein--Gordon norm is obtained
by multiplying the unit-spatial-norm mode by $(2E)^{-1/2}$.
The negative-frequency branch carries the opposite sign of the
Klein--Gordon norm, as required by the relativistic field inner product.
This separates the positive spatial localization density used to
display the radial bound states from the conserved Klein--Gordon
charge normalization \textcolor{revUnified}{\cite{BirrellDavies1982}}.
}

\section{Conclusion}
\label{sec:conclusion}

\setlength{\parindent}{0pt}

{We have developed a global spectral treatment of a charged Klein--Gordon field on the complete $(2+1)$-dimensional Ellis wormhole in a prescribed external magnetic field. The catenoidal embedding is used to define the ambient gauge connection; the scalar field and its dynamics remain intrinsic to the wormhole surface. Pullback produces a smooth azimuthal connection and a regular field-strength two-form. The associated oriented intrinsic magnetic pseudoscalar changes sign between the two asymptotic ends because the projection of the ambient field onto the continuously oriented surface normal changes sign. This is a geometrical projection effect, not a singularity or discontinuity of the electromagnetic field. The throat holonomy provides a complementary global gauge-invariant phase: it can be nontrivial even though the local magnetic pseudoscalar vanishes at the throat.

The radial Klein--Gordon amplitude and the one-dimensional Schr\"odinger wavefunction are related by the unitary transformation $u_\ell=\sqrt{r(x)}R_\ell$, with inverse $R_\ell=u_\ell/\sqrt{r(x)}$. Consequently, the original radial norm is $\int r|R_\ell|^2dx$, whereas the transformed norm is $\int|u_\ell|^2dx$. In the transformed representation the problem is a smooth, real, even Schr\"odinger operator on the complete line. For every $\alpha_B>0$, its effective potential grows as $\alpha_B^2x^2$ at both ends. The resulting confining operator has a discrete spectrum without imposing any boundary at the throat. This conclusion concerns the complete operator and does not depend on whether an eigenfunction is a finite Heun polynomial.

The confluent-Heun reduction identifies special parameter loci on which the Frobenius series terminates. Polynomial solvability requires both the spectral termination condition and the independent coefficient-compatibility condition; satisfying only the former is insufficient. The admissible families established here include degree-zero Gaussian modes, both global parity branches at degree one, and representative degree-two even states at integer azimuthal quantum numbers. Both positive algebraic roots are retained whenever they satisfy the stated admissibility requirements. These algebraic modes form exact states within the generally non-polynomial confined spectrum.

The full-line numerical calculation provides an independent check of this classification. Fourth-order finite differences, mesh refinement, interval variation, Richardson extrapolation, reflection-parity checks, and node counting reproduce the analytical benchmark levels and establish their positions in the global Sturm ordering. In particular, the odd degree-one polynomial benchmark is the lowest odd state at its parameter point, not the ground state of the full operator; the even nodeless state lies below it. The reported Richardson discrepancies quantify agreement for the specified discretization and should not be read as exact-arithmetic error bounds. The degree-two checks extend this comparison to additional integer angular momenta.

The normalized modes clarify the distinction between a radial Klein--Gordon amplitude, a positive spatial localization density, and the conserved Klein--Gordon charge norm. Exact spatial normalization of the degree-zero and degree-one modes uses the Ellis measure $r(x)dx$ and the corresponding Tricomi moments. Because the density is even, a unit-normalized mode has equal integrated spatial weight in the two asymptotic ends. The equal weights belong to a single global eigenstate defined on the connected wormhole. In the double-well regime, the throat barrier permits a useful symmetric/antisymmetric interpretation of the lowest even/odd pair as hybridized mouth-localized configurations. The forbidden-region action identifies the leading exponential scale of their splitting, but no tunnelling prefactor or dynamical transfer rate is claimed.

The magnetic field affects the radial shape through $|q\mathcal B_0|$ and the squared azimuthal number, while the sign-sensitive orbital coupling shifts the relativistic energy branches through $\sigma_B\ell$. Thus the field magnitude controls confinement, whereas the relative orientation of charge-field coupling and angular momentum controls the energy offset. The distinction between these effects, the throat holonomy, and the geometry-induced contribution to the effective potential gives a physically transparent interpretation of the spectrum. The present construction is a prescribed-background model; solving Maxwell's equations with dynamical sources on the wormhole, studying backreaction, or calculating a time-dependent inter-mouth transfer probability lies beyond its scope.

Our results connect the intrinsic geometry of the Ellis wormhole, an ambient-field-induced gauge connection, full-line self-adjoint spectral theory, conditional Heun-polynomial solvability, exact mode normalization, and independent numerical verification. The principal result is the identification of exceptional analytic modes and their spectral ordering within the complete magnetically confined spectrum.
}


\section*{Acknowledgements}

{The authors sincerely thank the Editor for overseeing the review process and for the opportunity to revise and improve the manuscript. We are particularly grateful to the Referee for the careful and thoughtful assessment of our work and for the valuable constructive comments and suggestions, which have helped us clarify the presentation and strengthen the manuscript.}

\section*{Credit authorship contribution statement}

\textbf{Abdullah Guvendi}: Conceptualization, Methodology, Formal Analysis, Writing - Original Draft, Investigation, Visualization, Writing - Review and Editing.\\
\textbf{Omar Mustafa}: Conceptualization, Methodology, Formal Analysis, Writing - Original Draft, Investigation, Visualization, Writing - Review and Editing.\\

\section*{Data availability}

All numerical data reported in this article are presented in the tables and figures.

\section*{Conflicts of interest statement}

The authors declare no competing interests.

\section*{Funding}

No external funding was received for this study.

\end{document}